\documentclass{aa}
\usepackage{graphicx}
\usepackage{txfonts}
\usepackage[numbers]{natbib}
\begin{document}

%\draft
\def\be{\begin{equation}}
\def\ee{\end{equation}}
\def\bea{\begin{eqnarray}}
\def\eea{\end{eqnarray}}
\def\c{\cite}
\def\nn{\nonumber}
\def\mcr{{{\rm M_{cr}}}}
\def\xo{{X_{o}}}
\def\dm{\Delta {\rm M}}
\def\ms{{\rm M_{\odot}}}
\def\mb{m_{B}}
\def\bo{B_{o}}
\def\cxo{C(x_{o})}
\def\rsix{R_{6}}
\def\vr{v_{{\rm r}}}
\def\vro{v_{{\rm ro}}}
\def\vt{v_{\theta}}
\def\po{\ifmmode P_{o} \else $P_{o}$ \fi}

\def\et{ {\it et al.}}
\def\la{ \langle}
\def\ra{ \rangle}
\def\ov{ \over}
\def\ep{ \epsilon}

\def\ep{\epsilon}
\def\th{\theta}
\def\ga{\gamma}
\def\Ga{\Gamma}
\def\la{\lambda}
\def\si{\sigma}
\def\al{\alpha}
\def\pa{\partial}
\def\de{\delta}
\def\De{\Delta}
\def\rsr{{r_{s}\over r}}
\def\rmo{{\rm R_{M0}}}
\def\rrm{{R_{{\rm M}}}}
\def\rra{{R_{{\rm A}}}}

\def\mdot{\ifmmode \dot M \else $\dot M$\fi}    % accretion rate
\def\mxd{\ifmmode \dot {M}_{x} \else $\dot {M}_{x}$\fi}
\def\med{\ifmmode \dot {M}_{Edd} \else $\dot {M}_{Edd}$\fi}
\def\bff{\ifmmode B_{{\rm f}} \else $B_{{\rm f}}$\fi}

\def\apj{\ifmmode ApJ \else ApJ \fi}    % lower
\def\apjl{\ifmmode  ApJ \else ApJ \fi}    %
\def\aap{\ifmmode A\&A \else A\&A\fi}    %
\def\mnras{\ifmmode MNRAS \else MNRAS \fi}    %
\def\nat{\ifmmode Nature \else Nature \fi}

\def\ms{\ifmmode M_{\odot} \else $M_{\odot}$\fi}    % lower
\def\na{\ifmmode \nu_{A} \else $\nu_{A}$\fi}    % Alfven frequency
\def\nk{\ifmmode \nu_{K} \else $\nu_{K}$\fi}    % Keplerian frequency
\def\ns{\ifmmode \nu_{{\rm s}} \else $\nu_{{\rm s}}$\fi}
\def\no{\ifmmode \nu_{1} \else $\nu_{1}$\fi}    % lower
\def\nt{\ifmmode \nu_{2} \else $\nu_{2}$\fi}    % upper
\def\ntk{\ifmmode \nu_{2k} \else $\nu_{2k}$\fi}    % upper
\def\dnmax{\ifmmode \Delta \nu_{max} \else $\Delta \nu_{2max}$\fi}
\def\ntmax{\ifmmode \nu_{2max} \else $\nu_{2max}$\fi}    % upper
\def\nomax{\ifmmode \nu_{1max} \else $\nu_{1max}$\fi}    % upper

\def\nh{\ifmmode \nu_{\rm HBO} \else $\nu_{\rm HBO}$\fi}    % HBO
\def\nqpo{\ifmmode \nu_{QPO} \else $\nu_{QPO}$\fi}    % HBO
\def\nz{\ifmmode \nu_{o} \else $\nu_{o}$\fi}    % HBO
\def\nht{\ifmmode \nu_{H2} \else $\nu_{H2}$\fi}    % HBO
\def\ns{\ifmmode \nu_{s} \else $\nu_{s}$\fi}    % stellar
\def\nb{\ifmmode \nu_{{\rm burst}} \else $\nu_{{\rm burst}}$\fi}
\def\nkm{\ifmmode \nu_{km} \else $\nu_{km}$\fi}    % stellar
\def\ka{\ifmmode \kappa \else \kappa\fi}    % stellar
\def\dn{\ifmmode \Delta\nu \else \Delta\nu\fi}    % stellar

\def\rs{\ifmmode {R_{s}} \else $R_{s}$\fi}    % stellar
\def\ra{\ifmmode R_{A} \else $R_{A}$\fi}    % Alfven radius
\def\rso{\ifmmode R_{S1} \else $R_{S1}$\fi}    % sonic point radius
\def\rst{\ifmmode R_{S2} \else $R_{S2}$\fi}    % sonic point radius
\def\rmm{\ifmmode R_{M} \else $R_{M}$\fi}    % stellar
\def\rco{\ifmmode R_{co} \else $R_{co}$\fi}    % stellar
\def\ris{\ifmmode {R}_{{\rm ISCO}} \else $ {\rm R}_{{\rm ISCO}} $\fi}
\def\rsix{\ifmmode {R_{6}} \else $R_{6}$\fi}
\def\rinfty{\ifmmode {R_{\infty}} \else $R_{\infty}$\fi}
\def\rinfsix{\ifmmode {R_{\infty6}} \else $R_{\infty6}$\fi}

\def\rxj{\ifmmode {RX J1856.5-3754} \else RX J1856.5-3754\fi}
\def\1739{\ifmmode {XTE J1739-285} \else XTE J1739-285\fi}
\def\exo{\ifmmode {EXO 0748-676} \else EXO 0748-676\fi}

\title{Testing the Accretion-induced Field-decay and Spin-up Model for Recycled Pulsars}
\titlerunning{Testing the Accretion-induced Field-decay and Spin-up Model}

\author{J. Wang
          \inst{1}
          \and
          C. M. Zhang
          \inst{2}
          \and
          H.-K. Chang
          \inst{1,3}
          }
%\authorrunning{Wang, Zhang \& Chang}
\institute{Institute of Astronomy, National Tsing Hua University,
Hsinchu 30013, Taiwan\\
\email{jwang@mx.nthu.edu.tw} \and National Astronomical
Observatories, Chinese Academy
of Sciences, Beijing 100012, China \\
\and Department of Physics, National Tsing Hua University, Hsinchu
30013, Taiwan}

\abstract {Millisecond radio pulsars have long been proposed to form from
a spin-up recycling process in a binary system. In this paper
we demonstrate that the  accretion-induced field-decay and
spin-up model for recycled pulsars can indeed produce those
millisecond pulsars with relatively weak magnetic fields of $10^8-10^9$ G and
short spin periods of a few milliseconds.
Our results also suggest that the value of the
currently observed highest spin frequency
of millisecond pulsars may simply be constrained by the amount of mass
available for accretion.
}
\maketitle

\section{Introduction}

\begin{figure}
\includegraphics[width=8cm]{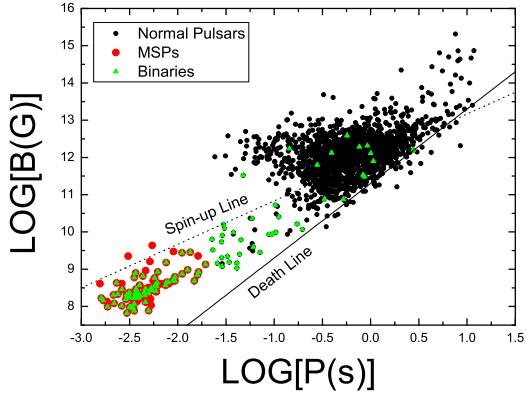}
\caption{Magnetic fields and spin periods of observed
pulsars (data taken from the ATNF pulsar catalogue). Black dots
are normal pulsars. Red dots represent MSPs.
Green triangles are pulsars in binaries. The "spin-up
line" represents the minimum spin-period to which a spin-up process may
proceed in an Eddington-limited accretion, while the "death-line"
corresponds to a polar cap voltage below which the pulsar activity
is likely to switch off (Bhattacharya \& van den Heuvel
1991).}\label{fig:B-P-ob}
\end{figure}

Radio pulsars are found to be of two kinds,
i.e., normal pulsars,
with magnetic field $B \sim 10^{12}$ G and spin period $P \sim$ a few seconds,
and millisecond pulsars (MSPs), which have low magnetic fields
($B \sim 10^{8}-10^{9}$ G) and short spin periods ($P \leq 20$ ms, e.g.
Bhattacharya \& van den Heuvel 1991; Lorimer 2008). These
bimodal distributions
in magnetic fields and spin periods
are shown, e.g., in
Fig. 1
in Wang et al. (2011) and also in
Fig.\ref{fig:B-P-ob} of this paper, which,
with data taken from ATNF
pulsar catalogue, shows a large population of
normal pulsars and a smaller population of MSPs.
These two populations are connected with a thin
bridge of pulsars in binaries.
Moreover, most MSPs are in
binary systems. It has long been proposed that
MSPs
are formed through a recycling process in which
neutron stars accrete material from their low-mass
companions  and are spun up by the angular momentum carried by the
accreted material. During this phase, the magnetic field is buried
by the accreted material and decays
(Alpar et al. 1982; Taam \& van den Heuvel 1986;
Bhattacharya \& van den Heuvel 1991; Radhakrishnan \& Srinivasan
1982; Bhattacharya \& Srinivasan 1995; van den Heuvel 2004).
These objects are therefore called recycled MSPs (Taam \& van den
Heuvel 1986; Bhattacharya \& van den Heuvel 1991; van den Heuvel
2004).

%%%%%%%%%Figure%%%%%%%%%%%%%%%%%%

%%%%%%%%%Figure%%%%%%%%%%%%%%%%%%

The evolutionary precursors of recycled MSPs are widely believed
to be pulsars in binaries with high magnetic field ($B \sim
10^{11}-10^{13}$ G) and long spin period ($P \sim$ a few seconds, Bhattacharya \&
van den Heuvel 1991; Liu, van Paradijs \& van den Heuvel 2007;
Lorimer 2010). It is evident that $B$ and $P$ of normal pulsars and
recycled pulsars are correlated with the duration of accretion
phase and the total amount of accreted mass (Taam \& van den Heuvel
1986; Shibazaki et al. 1989; Wijers 1997). If a neutron star accretes a small
amount of mass from its companion, e.g. $\sim 0.001 M_{\odot} -
0.01 M_{\odot}$, a recycled pulsar with mildly weak field and short
spin period ($B \sim 10^{10}$ G, $P \sim 50$ ms) will be formed
(e.g. Francischelli, Wijers \& Brown 2002), like PSR 1913+16 and PSR
J0737-3039 (Lyne et al. 2004; Kramer 2006).
Strong supporting evidences for
this recycling idea have been found in low mass X-ray binaries (LMXBs)
containing accreting millisecond X-ray pulsars (AMXPs), e.g. SAX J
1808.4-3658 (Wijnands \& van der Klis 1998), and in observing the
transition link from an X-ray binary to a radio pulsar PSR
J1023+0038 (Archibald et al. 2009).
At the end of the accretion
phase (accreted mass greater than $0.2 M_{\odot}$), the
neutron star magnetic field may arrive
at a bottom value of about $10^{8}-10^{9}$ G
and its spin period may reach a minimum of about a few milliseconds.
A millisecond pulsar is formed
(van den Heuvel \& Bitzaraki 1995ab; Urpin, Geppert
\& Konenkov 1998). The accretion-induced field-decay and spin-up
torque make the magnetic field and spin period (Hobbs \& Manchester
2004; Manchester et al. 2005) to change from $B \sim
10^{11}-10^{13}$ G and $P \sim$ a few seconds to $B \sim 10^{8}-10^{9}$ G and
$P \sim$ a few milliseconds. It is the recycling process that leads to the
bimodal distribution  of radio
pulsars.

In this paper, we examine the accretion-induced field-decay
and spin-up model of Zhang \& Kojima (2006) for these recycled pulsars.
We investigate the
differences in model-predicted $B$ and $P$ distributions with
different initial conditions.
We also compare those distributions with currently observed ones.
This paper
is organized as the following. Section 2 gives an overview of the model.
We describe the input parameters necessary in the calculations
of field and spin evolutions and analyze the results in Section 3.
Section 4 contains discussions and summary.

\section{The model}
\label{model}

\subsection{The recycling process}

A neutron star in a binary, with an initial magnetic field of about $B_0 \sim
10^{12}$ G and an initial spin period of about $P_0 \sim$ a few
seconds, can accrete material from its companion and forms an
accretion disk. When the ram pressure of accretion material equals
the magnetic pressure, a magnetosphere forms. As a result, a
boundary layer appears between the innermost disk and magnetosphere
due to the transition for rotating velocity of plasma from Keplerian
to the spin velocity of the neutron star (Inogamov \& Sunyaev 1999). In this
layer, the accreted matter will be channeled onto the polar patches
by the field lines, where the compressed accreted matter causes the
expansion of magnetic polar zone in two directions, downward and
equatorward (Zhang \& Kojima 2006). Therefore, the magnetic flux in
the polar zone is diluted, and more matter is accreted to the polar
cap and diffuses to the surface of the neutron star. Finally, the polar cap
area expands and occupies the entire neutron stars surface, and the magnetic
flux is buried in the equatorial area. The magnetosphere is then
compressed to the neutron star surface,
leading to an object with weak fields
in large scale (about $\sim 10^8$ G) and very strong fields in small
scale (about $\sim 10^{14}$ G). Meanwhile, the angular
momentum carried by the accreted matter spins up the neutron star,
forming a MSP with the spin period of a few milliseconds.

\subsection{Magnetic field evolution}

Based on the above accretion-induced field-decay and spin-up model,
the accretion-induced field and spin evolution is obtained
analytically: (Zhang \& Kojima
2006)
\begin{equation}
\label{bt} B(t) = \frac{B_f}{(1 - [C\,e^{-y}-1]^2)^{\frac{7}{4}}}\,\,\, ,
\end{equation}
where we have $y = \frac{2 \Delta
M}{7M_{cr}}$, the accreted mass $\Delta M = \dot{M} t$, the crust
mass $M_{cr} \sim 0.2 M_\odot$, and $C = 1 + \sqrt{1-x_0^2} \sim 2$ with
$x_0^{2} = (\frac{B_f}{B_0})^{4/7}$. $B_0=B(t=0)$ is the initial field
strength and $B_f$ is the bottom magnetic
field, which is defined by the neutron star magnetosphere radius matching the
stellar radius, i.e., $R_M(B_f) = R$. $\rrm$ is defined as $R_M =
\phi R_A$ where Alfv\'{e}n radius $R_A =3.2\times 10^8 {\rm cm}
\dot{M}^{-2/7}_{17} \mu^{4/7}_{30} m^{-1/7}$ (Elsner \& Lamb 1977;
Ghosh \& Lamb 1977).
%\nn
% &=& 0.82\times 10^6 ({\rm cm})
%\dot{M}^{-2/7}_{18} B^{4/7}_{8} m^{-1/7}\rsix^{12/7}\;. \label{ra}
%\eea
The model dependent parameter $\phi$ is about 0.5 (Ghosh \& Lamb
1979b; Shapiro \& Teukolsky 1983; Frank et al. 2002).
$\dot{M}_{17}$ is the accretion rate in units of $10^{17}$ g/s. $\mu
_{30}$ is the magnetic moment in units of $10^{30} {~} {\rm
G~cm^3}$. The mass $m = M/M_{\odot}$ is in the unit of solar mass.
According to this model, the bottom field of neutron stars is determined by the
condition that the magnetosphere radius equals the neutron stars radius (Zhang
\& Kojima 2006). Using the relation $R_M(B_f) = R$, we can obtain
the bottom field,
\begin{equation}
B_f=1.32\times10^8{\rm G}(\frac{\dot{M}}{\dot{M}_{18}})^{\frac{1}{2}}
m^{\frac{1}{4}}R_6^{-\frac{5}{4}}\phi^{-\frac{7}{4}},
\end{equation}
where $\dot{M}_{18} = \dot{M}/10^{18}$ g/s and $\rsix = R/10^6$ cm.

\subsection{Spin evolution}

During the accretion phase, the neutron star is spun up by the angular
momentum carried by the accreted matter. The spin evolves according
to the following relation
given by Gosh \& Lamb (1979b):
\begin{eqnarray}
-\dot{P}&=&5.8 \times 10^{-5}[(\frac{M}{M_\odot})^{-\frac{3}{7}}
R^{\frac{12}{7}}_6 I_{45}^{-1}]\nonumber\\~& & \times
B^{\frac{2}{7}}_{12} (PL^{\frac{3}{7}}_{37})^2n(\omega_s) {\empty ~
~ }  s~yr^{-1},\label{pdot}
\end{eqnarray}
where we define the parameters, the surface field $B_{12} =
B/10^{12}$ G, the moment of inertia $I_{45} = I/10^{45} {\rm
g~cm^2}$, the X-ray brightness ($L = GM\dot{M}/R$) $L_{37}$ in units
of $10^{37}$ erg/s, respectively. The dimensionless parameter
$n(\omega_s)$ is the fastness parameter,  whose
expression is given by Gosh \& Lamb (1979b),
\begin{equation}
n(\omega_{s}) = 1.4\times \left(
\frac{1-\omega_{s}/\omega_{c}}{1-\omega_{s}} \right). \label{ns}
\end{equation}
where $\omega_s$ is defined as
\begin{equation}
\omega_s\equiv\frac{\Omega_s}{\Omega_k(R_M)} =
1.35[(\frac{M}{M_\odot})^{-2/7}R^{15/7}_6] B^{6/7}_{12} P^{-1}
L^{-3/7}_{37} \,\,\, , \label{omega}
\end{equation}
with $\Omega_s$ being the stellar spin frequency and $\Omega_k$ is the
Keplarian frequency.
$\omega_s$ is the ratio parameter of the angular velocities
which describes the relative importance of stellar rotation and
plays a significant role in our entire understanding of accretion to
the rotating magnetic neutron stars (Elsner \& Lamb 1977; Ghosh \& Lamb 1977;
Li \& Wang 1996, 1999; Shapiro \& Teukolsky 1983).
For a slowly rotating magnetic neutron star,
$\omega_s\ll 1$.
$\omega_c$ depends on several properties of the system (Gosh \& Lamb 1979b)
and is taken to be 0.35 in this computation.

\section{$B$ and $P$ distributions of MSPs}

\subsection{Input parameters}
\label{input}

According to the accretion-induced field-decay and spin-up model by
Zhang \& Kojima (2006), a pulsar with strong magnetic field (e.g. $B
\sim 10^{12}$ G) and slow rotation ($P \sim$ a few seconds) in a
binary system may be spun up to become a millisecond pulsar via accretion,
and the
magnetic field decays to a bottom value ($B \sim 10^{8}-10^{9}$ G) during
this phase. The final state of a recycled system is characterized by
the magnetic field and spin period, which are related to the
initial magnetic field, initial spin period, accretion rate,
accretion time, and the mass and radius of neutron stars.
In computing the model-prediction of $B$ and $P$,
we adopt these input parameters as
the following:
\newline
(1) All precursors to recycled MSPs are assumed to be normal
pulsars with magnetic field of about $B_0 \sim 10^{12}$ G and spin
period of about a few seconds. According to the data taken from ATNF
catalogue, the $\log B$ and $\log P$ distributions of normal pulsars
can be
well described by a Gaussian function (see Wang et al. 2011).
We therefore take lognormal distributions
as the inputs of initial magnetic fields and spin periods.
The probability density
function reads,
\begin{equation}
p(x)=\frac{1}{\sqrt{2\pi}\sigma}\exp(-\frac{1}{2}[\frac{x-\mu}{\sigma}]^2),\label{Gaussian}
\end{equation}
where $\mu$ is the mean value of the distribution, and
$\sigma$ is the standard deviation. We take $\mu =\log B({\rm G})=
12$  and $\sigma = 0.5$ for the $B$ distribution
(see e.g. Hartman et al. 1997; Hobbs et al.
2011; Kaspi 2010; Wang et al. 2011) and $\mu = \log P({\rm s}) =
0$ and $\sigma = 0.4$ for the $P$ distribution
(see e.g. Lorimer 2010 and references
therein). The range of $B$ and $P$ to consider is taken as
$B_0 = 10^{10.5}-10^{14.0}$ G and $P_0 = 0.1-30$ s, respectively.
\newline
(2) The accretion rate is taken in the range from $\dot{M} = 10^{16}$
g/s to $\dot{M} = 10^{18}$ g/s (see e.g. Wijers 1997)
and is assumed to be constant during the whole process.
The duration of the accretion phase is in the range
from $\Delta t = 10^{7}$ yr to $\Delta t = 10^{9}$ yr.
Most systems may accrete $0.1 - 0.2 M_{\odot}$ at the end of the
accretion phase (see Shapiro \& Teukolsky 1983). However, in some
binary systems, the maximum accretion mass can be $0.6 - 0.8
M_{\odot}$ (van den Heuvel 2011), depending on
the intrinsic properties of the system.
We adopt lognormal
distributions (see Eq. \ref{Gaussian}) for the accretion rate and
the accretion time. The mean values and standard deviations are $\mu
= \log \dot{M}({\rm g/s}) = 17$, $\sigma = 0.1$ for the accretion rate
and $\mu = \log\Delta t ({\rm yr}) = 8$,
$\sigma = 0.4$ for the accretion time respectively.
\newline
(3) {\bf According to recent statistics of neutron star mass (see
Zhang et al. 2011), we consider a Gaussian mass distribution
with the mean at $1.4 M_{\odot}$ and the standard deviation equal
to 0.2 $M_{\odot}$ within the range from 0.9 $M_{\odot}$ to 2.2 $M_{\odot}$.}
\newline
(4) It is widely believed that the neutron star radius is about 10 km.
We consider a uniform distribution of the radius from 10 km to 20 km.

\subsection{Distribution of recycled MSPs}

%%%%%%%%%Figure%%%%%%%%%%%%%%%%%%
\begin{figure}
\centering
\includegraphics[width=4.5cm]{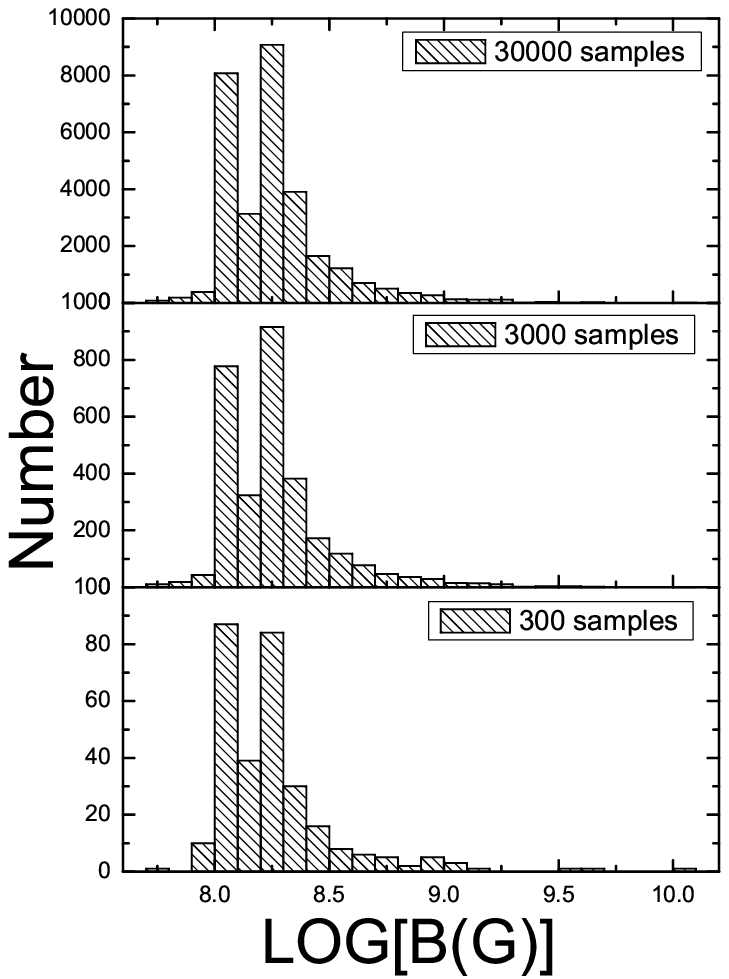}%
\includegraphics[width=4.5cm]{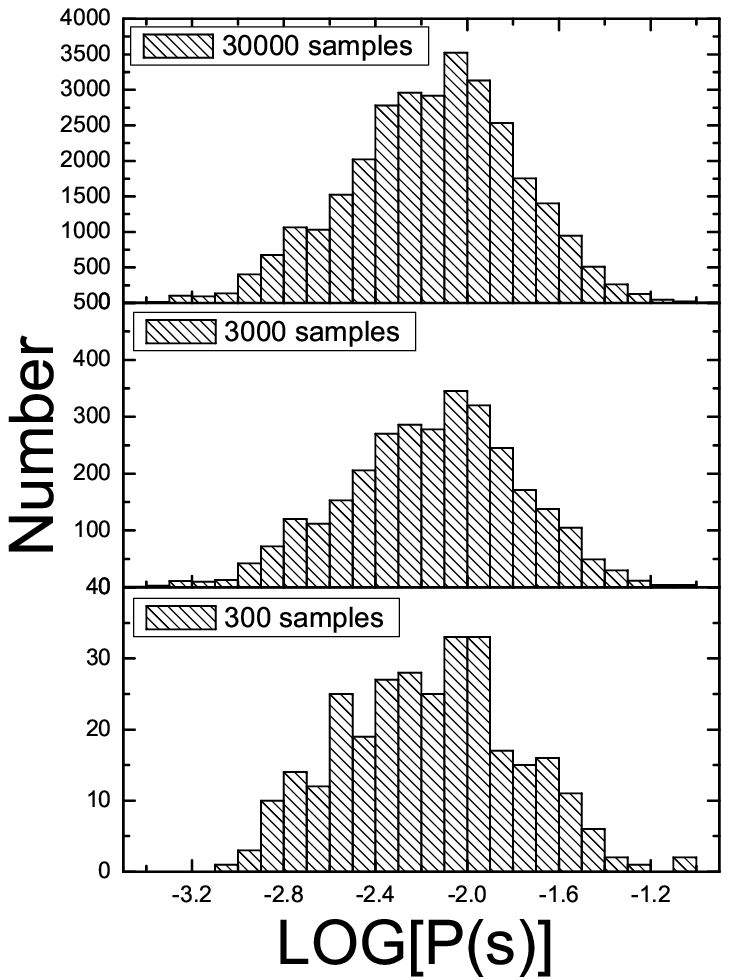}
\caption{The computed $B$ and $P$ distributions of recycled MSPs with
different numbers of random input samples. The left
(right) panel is the $B$ ($P$) distribution.
As discussed in the text, the computed distributions
with 30000 random samples are good enough and are the ones
used for comparison with observation in this work.}\label{fig:b-his-com}
\end{figure}
%%%%%%%%%Figure%%%%%%%%%%%%%%%%%%
%%%%%%%%%Figure%%%%%%%%%%%%%%%%%%
\begin{figure*}
\centering
\includegraphics[width=8.5cm]{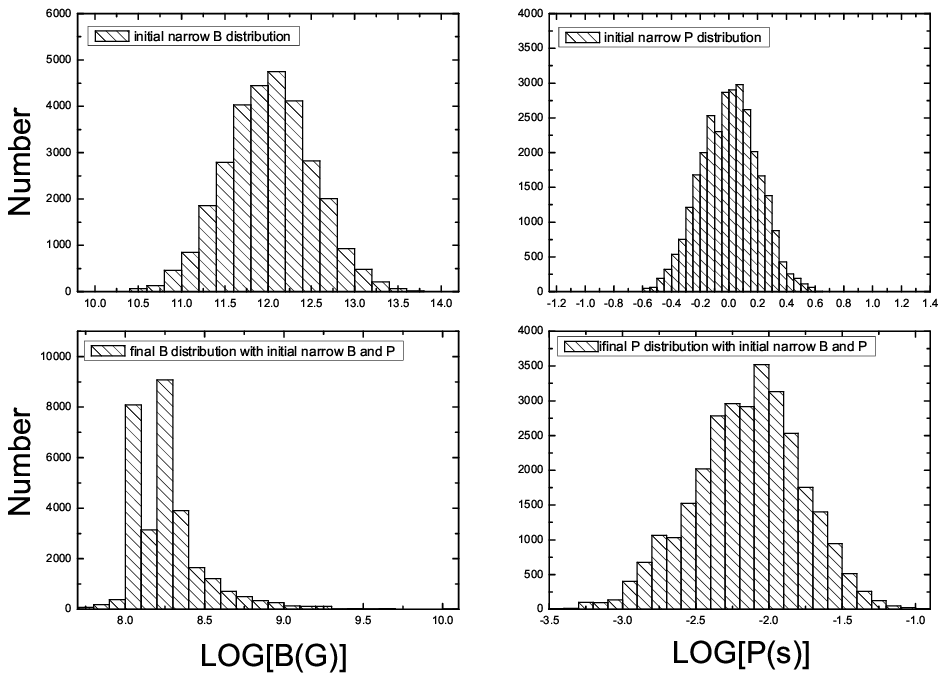}%
\includegraphics[width=8.5cm]{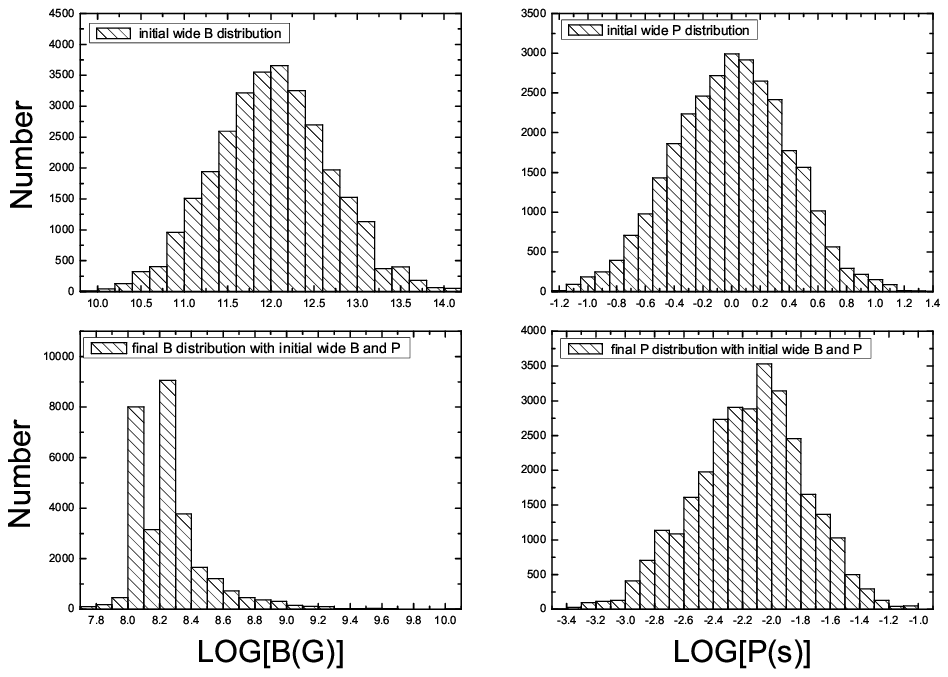}
\caption{$B$ and $P$ input distributions with narrow and wide
Gaussian widths (the upper panels) and their resultant
final $B$ and $P$ distributions (the lower panels).
}\label{fig:gau-bp-com}
\end{figure*}
%%%%%%%%%Figure%%%%%%%%%%%%%%%%%%

We calculated the final $B$ and $P$ according to the accretion-induced $B$ and
$P$ evolution model described in section \ref{model}.
In order to produce the model-predicted distribution,
we generated a large number of input parameter random samples from
the input parameter distributions mentioned  in section \ref{input}.
Different computation runs with 300, 3000 and 30000 samples were performed
to examine how many samples are needed to have a converged result.
Fig. \ref{fig:b-his-com} shows the $B$ and $P$ distributions
of recycled MSPs  obtained with 300, 3000 and 30000 random samples,
respectively.
We can see that, with the initial conditions considered in section \ref{input},
the Zhang \& Kojima (2006) model indeed expects
the
magnetic fields of recycled MSPs are of
$10^{8}-10^{9}$ G, and their spin periods are a few
milliseconds.
We divided these distributions into  histograms with bin size of 0.1
and compared their difference with the
Kolmogorov-Smirnov (K-S) test {\bf (Press et al. 1992)}.
For the comparison of $B$ distributions between 300 and 3000 samples,
the K-S statistic gives a 86.1\% probability of the two being drawn from the
same parent distribution.
That probability for the two distributions of 3000 and 30000 samples
is 100.0\%.
For the $P$
distributions, that probability is 64.9\% for that of 300 and 3000 samples, and
again reaches 100.0\% for that of 3000 and 30000 samples.
We therefore content ourselves with the results obtained with
30000 random samples.

Although the initial conditions, that is, input parameters, as
described in section \ref{input} are quite commonly adopted,
we further checked the sensitivity of the model prediction results
to the variation in the distribution of the input parameters.
We used  different input widths of the lognormal $B$ and $P$ distributions,
which are 0.375 (instead of 0.5) for $B$ and 0.2 (instead of 0.4) for $P$.
The ranges exploited for the $B$ and $P$ distributions are the same
as in
section \ref{input}.
These input distributions and the corresponding computed
final $B$ and $P$ distributions are
shown Fig. \ref{fig:gau-bp-com}.
Based on the K-S test, the similarity between the final distributions obtained
with narrow and wide inputs is 100.0\%, for both $B$ and $P$ distributions.
They are insensitive to mild variation of their progenitor distributions.

\begin{figure}
\centering
\includegraphics[width=8cm]{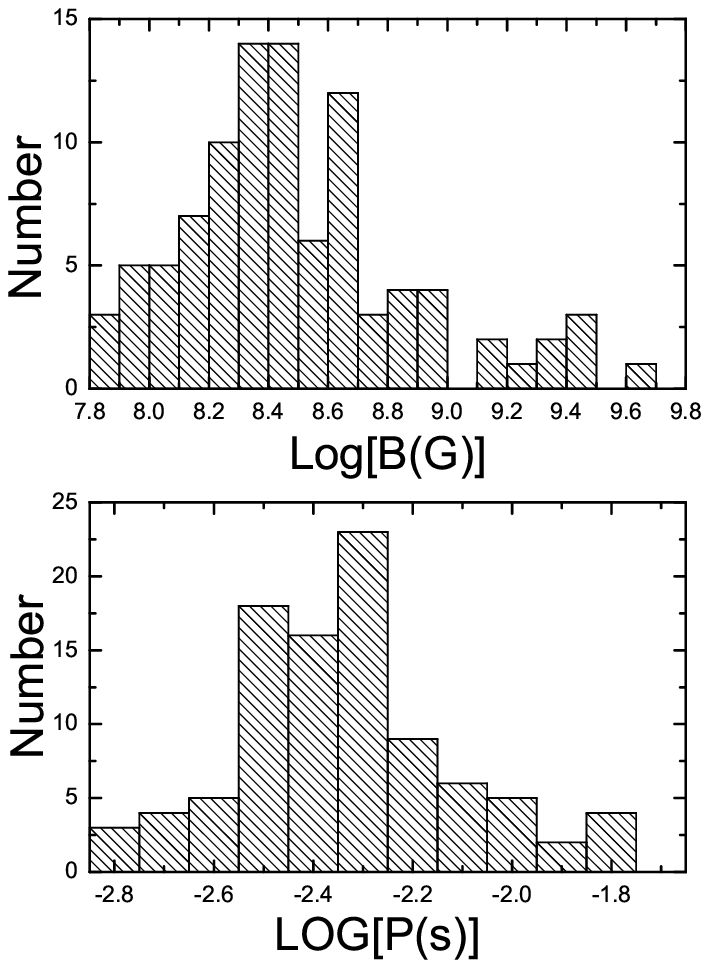}
\caption{The $B$ and $P$ distributions of observed MSPs (data taken
from the ATNF
pulsar catalogue).} \label{fig:ob-msp}
\end{figure}

\begin{figure}
\centering
\includegraphics[width=8cm]{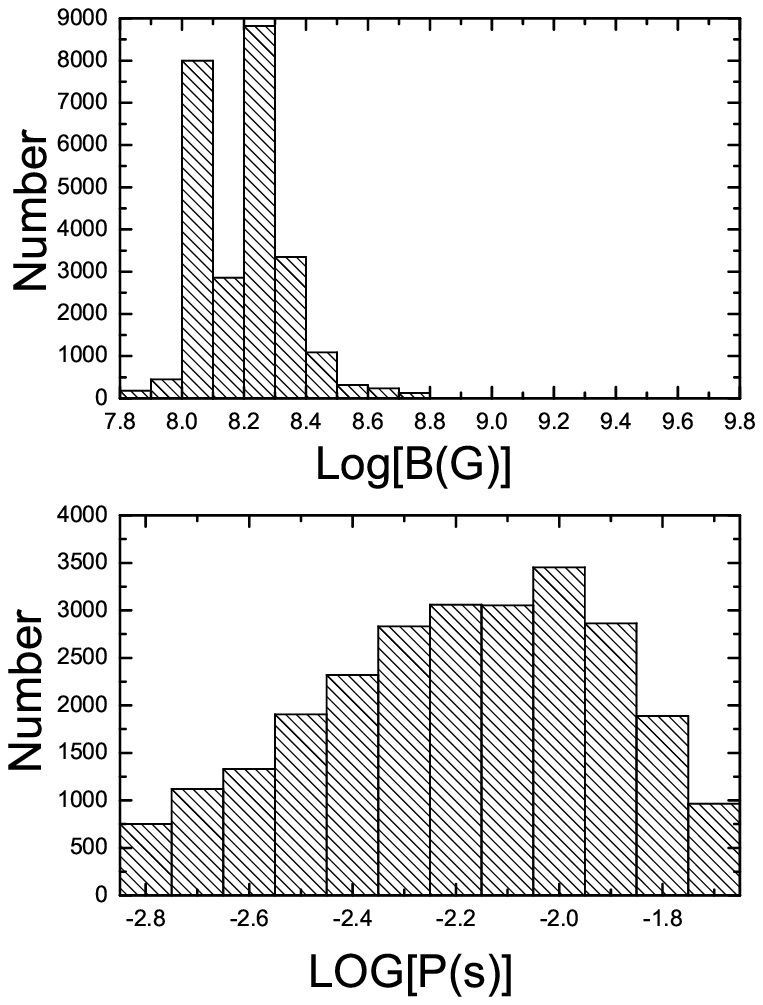}
\caption{The computed $B$ and $P$ distributions of recycled MSPs
with spin periods shorter than 20 ms.}\label{fig:forcom}
\end{figure}

Finally, we compared the model-predicted $B$ and $P$ distributions
with currently observed ones.
The observed MSPs are usually defined as pulsars with spin period
less than 20 ms. Fig. \ref{fig:ob-msp} shows the $B$ and $P$
distributions of observed MSPs.
Because of the wide ranges of the input parameters,
we have some MSPs with longer computed spin periods
(see Fig. \ref{fig:b-his-com}).
The comparison with observation is conducted
for spin periods shorter than 20 ms, and the computed
$B$ and $P$ distributions
for these recycled pulsars are plotted in Fig. \ref{fig:forcom}.
Again using the K-S test,
we found that the degree of
similarity for the $B$ distributions is $69.2\%$ and that of the $P$
distributions
is $73.6\%$.
We also performed a $\chi^2$ test to check the consistency between the
model predicted distributions and observed ones.
The errors in the observed distributions
 are assigned as $\sqrt{N_i}$ for each bin,
where $N_i$ is the number of observed recycled pulsars in the $i$th bin.
The $\chi^2$ value is $22.8$ (19 degrees of freedom) for the $B$ distribution,
which corresponds to a null hypothesis probability of 24.6\%,
 and
$16.6$ (11 degrees of freedom) for the $P$ distribution,
which corresponds to a 12.0\% null hypothesis probability.
Both the K-S test and $\chi^2$ test indicate that the
consistency between the currently observed distributions
and the Zhang \& Kojima (2006) model prediction
is roughly at about the so-called 1-$\sigma$ level.
In other words, although not of a high degree of similarity,
they are not inconsistent to each other.

%%%%%%%%%Figure%%%%%%%%%%%%%%%%%%
%%%%%%%%%Figure%%%%%%%%%%%%%%%%%%
%%%%%%%%%Figure%%%%%%%%%%%%%%%%%%
%%%%%%%%%Figure%%%%%%%%%%%%%%%%%%

\subsection{$B$ and $P$ evolutions of recycled MSPs}

\begin{figure*}
\centering
\includegraphics[width=8cm]{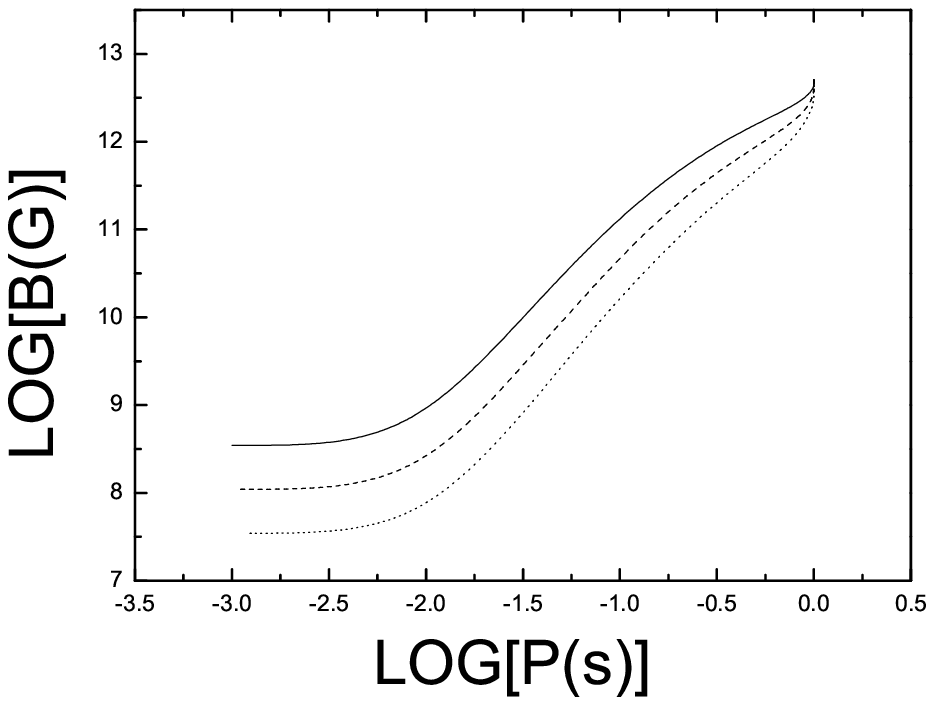}
\includegraphics[width=8cm]{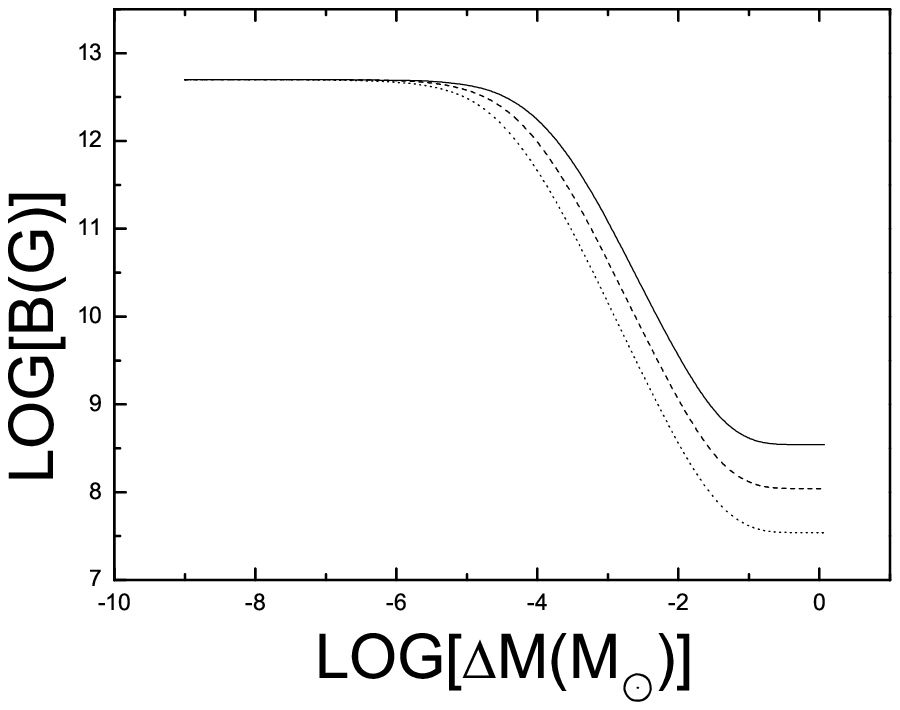}
\newline
\includegraphics[width=8cm]{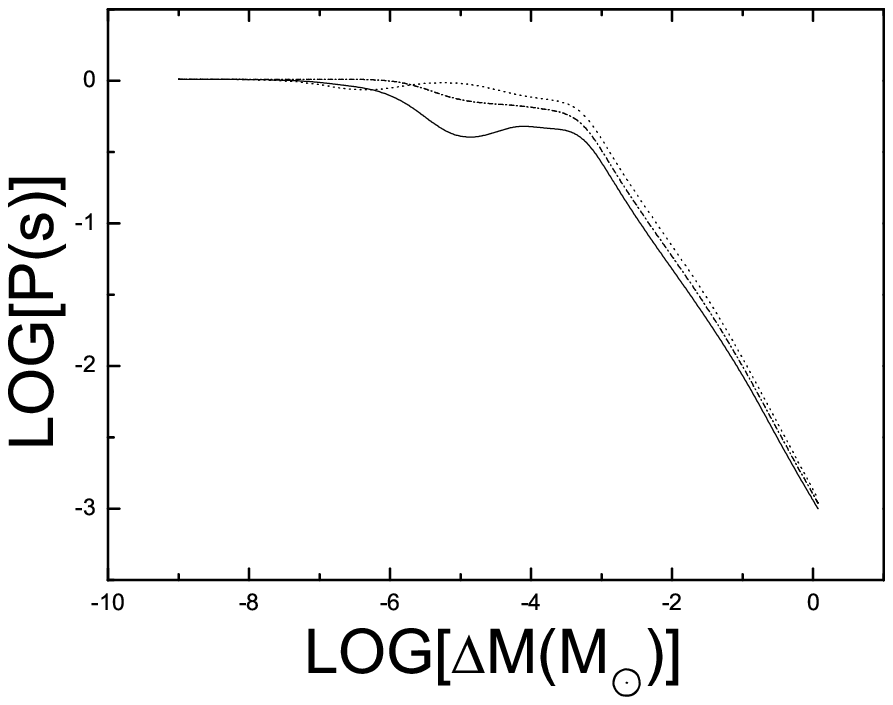}
\includegraphics[width=8cm]{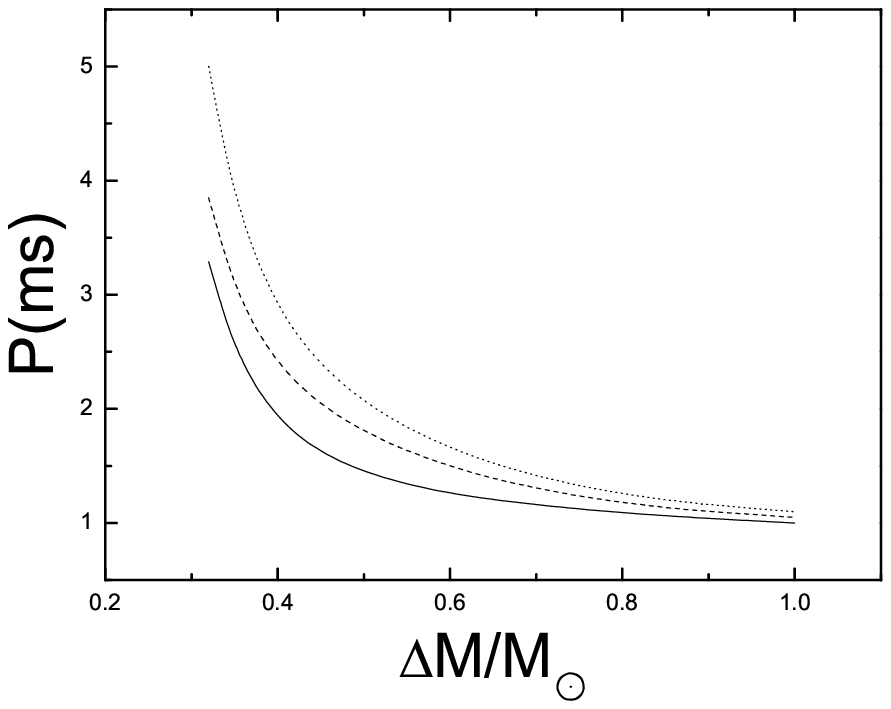}
\caption{$B$ and $P$ evolution in the recycling process.
The upper left panel shows the joint evolution of $B$ and $P$.
The upper right and lower left panels are their evolution as a function
of accreted mass
$\Delta M$. The solid, dashed,
and dotted lines are the evolutionary tracks
with the accretion rate of $10^{18}$ g/s, $10^{17}$ g/s,
and $10^{16}$ g/s, respectively. The initial $B$
and $P$ are taken as $B_0 = 5\times10^{12}$ G and $P_0 = 1$ s.
The lower right panel is a zoom-in view of the lower left panel in a linear
scale for spin periods shorter than 5 ms.}
\label{fig:B-P-deltam}
\end{figure*}

Fig. \ref{fig:B-P-deltam} displays the $B$ and $P$ evolutionary tracks
during the accretion phase with initially slowly rotating and highly
magnetized progenitors.
In this calculation, we took  the initial magnetic field
as $B_0 = 5\times10^{12}$ G and the initial spin period
as $P_0 = 1$ s.
Accretion rates of $10^{18}$ g/s, $10^{17}$ g/s, and $10^{16}$ g/s
were considered.
These tracks were followed until the accreted mass reached 1 $M_{\odot}$.
The upper right panel in Fig. \ref{fig:B-P-deltam}
shows that the magnetic field
decays with the accumulation of the accreted material and that the
bottom values of magnetic fields
correlate with the accretion rates.
The lower left panel shows that the pulsar rotates faster and faster
when accreting more and more mass and the spin
period is insensitive to the accretion rate after
accreting about $0.001 M_{\odot}$.

%%%%%%%%%Figure%%%%%%%%%%%%%%%%%%
%%%%%%%%%Figure%%%%%%%%%%%%%%%%%%

\section{Discussion and Summary}

We tested the accretion-induced field-decay and spin-up model for
recycled MSPs (Zhang \& Kojima 2006, Wang et al. 2011). In
our computation, we considered lognormal distributions of initial
$B$ and $P$
in the range of $B_0 = 10^{10.5}-10^{14}$ G
and $P_0 = 0.1-30$ s,
a Gaussian neutron star mass distribution in the range of $M = 0.9-2.2
M_{\odot}$, a uniform neutron star radius distribution in
$R = 10-20$ km,
, a lognormal distribution of
the accretion rate in $\dot{M} = 10^{16}-10^{18}$ g/s and of the
accretion time in $\Delta t
= 10^{7}-10^{9}$ yr.
We found that the computed $B$ and $P$ distributions of recycled MSPs
are insensitive to mild variations in the width of the
initial distributions.
Based on the K-S test and $\chi^2$ test, we found that
the Zhang \& Kojima (2006) model prediction is consistent with observation
at the 1-$\sigma$ level.

The accretion-induced field-decay model is based on the idealized
idea of dilution of polar magnetic flux due to accretion.
All the possible instabilities are ignored.
Besides, a constant accretion rate is assumed
during the whole accretion process. There are, however, some
specialities for each system, such as the influence of thermal and
viscous instabilities of the accretion disk and the propeller effect on
the mass transfer process.
The orbital angular momentum loss of the
system and its causes may also have some effect on the final $B$ and $P$
state. The numerous plasma instabilities, such as the
Rayleigh-Taylor instability and the Kelvin-Helmholtz instability
(Ghosh \& Lamb 1979a), may result in the penetration of the
magnetosphere, prying the field lines aside and azimuthally wrapping
the field lines by the disk matter (e.g. Romanova 2008; Kulkarni \&
Romanova et al. 2008), which in turn may modify the field strength
evolution and then perturb the spin evolution.
During the accretion, the accretion rate may change
due to changes of the system and some instabilities,
which can lead to the difference between
the theoretical results and the actual values.

It is generally believed that there are two possible ways to
form MSPs,
i.e. the standard accretion-induced field-decay and spin-up model
and accretion-induced collapse of white dwarfs (AIC).
Due to the conservation of magnetic flux during the collapse
of white dwarfs, the MSPs formed via AIC are expected
to have high magnetic fields and short spin periods.
Although the number of MSPs formed via AIC is no more than 20\%
(Zhang et al. 2011), this may also contribute to make
the observed $B$ and $P$ distributions of MSPs somewhat different
from that of MSPs formed via only the standard mechanism.

Some selection effects should also be noted when comparing the model prediction
with observation. For example,
the spin-down energy loss rate, $\dot{E}\propto\frac{B^2}{P^4}$,
is related to the radio power of a pulsar.
MSPs with relatively stronger magnetic fields and shorter spin periods
are easier to be observed.

Owing to the ohmic dissipation,
the buried field may re-emerge after accretion
(e.g. Young \& Chanmugam 1995; Bhattacharya 2008). Besides, If the
magnetic field of recycled pulsars arrive at the bottom value at
this stage and the accretion has not yet ended, the accumulation of
material may spin up the recycled pulsar further. However, the
magnetic field of neutron stars will not decay.
The magnetic flux carried by the
plasma accreted onto the neutron star surface may increase
the neutron star surface field strength.

We also plot the $B$ and $P$ evolutionary scenarios during the accretion
process in Fig. \ref{fig:B-P-deltam}. All of these plots show
that the bottom
magnetic field strength is different for different accretion rates
and that the minimum period is insensitive to the accretion rate at
the end of the accretion phase.
After accreted 1 $M_{\odot}$, the spin period reaches about 1 ms.
If the there is enough mass that can be accreted (e.g. 1.2 $M_{\odot}$),
the spin period can be shorter than one millisecond,
forming a submillisecond pulsar.
It is claimed that the maximum accreted mass is 0.8 $M_{\odot}$
under the assumption of 1 $M_{\odot}$ companion star
{van den Heuvel \& Bitzaraki 1995a; Wijers 1997}.
From the recent statistics of neutron stars/LMXB
(Liu, van Paradijs \& van den Heuvel 2007),
most neutron stars/LMXBs have the companion mass of 0.7 $M_{\odot}$.
According to the accretion-induced field-decay and spin-up model,
the spin period can reach the value of 1.1-1.2 ms
after accreting 0.7-0.8 $M_{\odot}$.
So far, the observed shortest period for MSPs is 1.4 ms (Hessels et al. 2006),
and that for millisecond X-ray pulsars is 1.6 ms (e.g. Patruno 2010).
Those two spin frequencies, 716 Hz and 620 Hz, are both lower
than than the believed break-up
spin frequency $\sim$ 1000 Hz (e.g. Lattimer \& Prakash 2004).
Models to explain the lack of submillisecond pulsars have been proposed,
including angular momentum loss due to gravitational radiation
and the magnetic spin equilibrium.
A significant quadrupole moment may exist  due to some oscillation modes
or the so-called 'crustal mountains' and 'magnetic deformation', which
leads to the emission of gravitational waves and the loss of angular momentum,
in particularly because of its strong dependence on the spin frequency to the
5th power
(e.g. Bildsten 1998; Haskell \& Patruno 2011;
Patruno, Haskell \& D'Angelo 2011).
On the other hand, however, the magnetic braking resulting
from the action of stellar wind indirectly carries
away the angular momentum (Rappaport, Verbunt \& Joss 1983),
which is more efficient than gravitational
radiation by about 2 orders of magnitude
in some close systems (Kalogera, Kolb \& King 1998).
Furthermore, the magnetic spin equilibrium set by disk/magnetosphere coupling
seems successful in explaining the lack of submillisecond pulsars
(D'Angelo \& Spruit 2011; Kajava et al. 2011).
In supplement to those efforts, the results
of the accretion-induced field-decay and spin-up model as presented in this
paper suggest that the achievable minimum spin period due
to the recycling process may depend mainly on the amount of mass available
for accretion, before the limiting shortest period
set by the magnetic spin equilibrium is reached.

\section*{acknowledgements}

We appreciate very much the valuable comments from anonymous referees,
 which improved this paper a lot.
This work was partially supported by the National Natural Science Foundation of
China (NSFC 10773017) and the National Basic Research Program of
China (2009CB824800). It was also supported by the National Science Council
of Taiwan under grant NSC 99-2112-M-007-017-MY3.


\begin{thebibliography}{99}

\bibitem{}
Alpar M. A., Cheng A. F. \&  Ruderman M. A. et al. 1982, Nature,
300, 728
%A new class of radio pulsars

\bibitem{}
Archibald A. M., Stairs I. H. \& Ransom S. M. et al. 2009, Science,
324, 1411
%A Radio Pulsar/X-ray Binary Link

\bibitem{}
Bhattacharya D. \& van den Heuvel E. P. J. 1991, Phys. Rep., 203, 1

\bibitem{}
Bhattacharya D. \& Srinivasan G. 1995, in X-ray Binaries, eds. Lewin
W. H. G., van Paradijs J. and van den Heuvel E. P. J., (Cambridge
University Press)

\bibitem{}
Bhattacharya D. 2008, AIPC, 1068, 137

\bibitem{}
Bildsten L. 1998, ApJ., 501, L89%Gravitational Radiation and Rotation
%of Accreting Neutron Stars

%\bibitem{}  Camilo F., Thorsett S. E. \& Kulkarni S. R. 1994, ApJ., 421, L15
%The magnetic fields, ages, and original spin periods of
%millisecond pulsars

\bibitem{}
D'Angelo, C. R., \& Spruit, H. C. 2011, MNRAS, 416, 893

\bibitem{}
Elsner R. F., \& Lamb F. K. 1977, ApJ., 215, 897

\bibitem{}
Francischelli G. J., Wijers R. A. M. J. \& Brown G. E. 2002, ApJ,
565, 471
%The evolution of relativistic binary progenitor systems.
%Relativistic binary pulsars, such as B1534+12 and B1913+16,
%having close orbits with a binary separation of ~3 R¨'.

\bibitem{}
Frank J., King A. \& Raine D. J. 2002, Accretion Power in
Astrophysics, Cambridge, UK

\bibitem{} Ghosh P. \& Lamb F.K. 1977, ApJ., 217, 578

\bibitem{} Ghosh P. \& Lamb F.K. 1979a, ApJ., 232, 259

\bibitem{} Ghosh P. \& Lamb F.K. 1979b, ApJ., 234, 296

\bibitem{}  Hartman J.W. et al. 1997, 325, 1031

\bibitem{}
Haskell, B., \& Patruno, A. 2011, ApJL, 738, L14

\bibitem{}
Hessels J. W., Ransom S. M. \& Stairs I. H. et al.
%Freire, P. C. C.,
%Kaspi, V. M., Camilo, F.
2006, Science, 311, 1901
%716 Hz pulsar

\bibitem{}
Hobbs G. \& Manchester R. 2004, ATNF Pulsar Catalogue,
http://www.atnf.csiro.au/research/pulsar/psrcat/psrcat\_help.html

\bibitem{}
Hobbs G. et al. 2011, PASP, 28, 202

\bibitem{}
Inogamov N. A. \& Sunyaev R. A., 1999, AstL., 25, 269

\bibitem{}
Kajava, J. J. E., Ibragimov, A., Annala, M., Patruno, A., \& Poutanen, J. 2011,
MNRAS, 417, 1454

\bibitem{}
Kaspi V. M. 2010, PNAS, 107, 7147, arXiv:1005.0876v1

%\bibitem{}
%Kiziltan B., Kottas A., Thorsett S. E. 2010,
%arXiv1011.4291[astro-ph]
%The Neutron Star Mass Distribution
\bibitem{}
Kalogera V., Kolb U. \& King A. R. 1998, Apj., 504, 967

\bibitem{}
Kulkarni A. K., \& Romanova M. M. 2008, MNRAS, 386, 673

\bibitem{}
Kramer M. et al. 2006, Science, 314, 97

\bibitem{}
Lattimer J. M. \& Prakash M. 2004, Science, 304, 536

\bibitem{} Li X. D., Wang Z. R., 1996, A\&A, 307, L5
%fastness

\bibitem{}
Li X. D. \& Wang Z. R. 1999, ApJ, 513, 845
%Disk Accretion onto Magnetized Neutron Stars: The Inner Disk Radius and Fastness Parameter

\bibitem{}
Liu Q. Z., van Paradijs J., van den Heuvel E. P. J. 2007, A\&A, 469,
807
%Catalogue of Galactic low-mass X-ray binaries

\bibitem{}
Lorimer D. R. 2008, Living Rev. Relativity, 11, 8
http://relativity.livingreviews.org/Articles/lrr-2008-8/
%arxiv:0811.0762

\bibitem{}
Lorimer D. R. 2010, arXiv:1008.1928[astro-ph]
%Radio pulsar populations

\bibitem{}
Lyne A. G., Burgay M., Kramer M. et al.
2004, Science, 303, 1153 %(astro-ph/0401086) A Double - pulsar system -
%A Rare laboratory for relativistic gravity and plasma physics.

\bibitem{}
Manchester R. N., Hobbs G. B., Teoh A. \& Hobbs M. 2005, AJ,
129, 1993 %arxiv:astro-ph/0412641 The ATNF pulsar catalogue.

\bibitem{}
Shibazaki N., Murakami T., Shaham J. \& Nomoto K. 1989, Nature, 342,
656
%Does mass accretion lead to field decay in neutron stars

\bibitem{}
Shapiro S. L. \& Teukolsky S. A. 1983, Black Holes, White Dwarfs and
Neutron Stars. Wiley, New York

\bibitem{}
Patruno, A., Haskell, B., \& D'Angelo, C. 2011, arXiv:1109.0536

\bibitem{}
Patruno, A. 2010, ApJ, 722, 909

{\bf \bibitem{}
Press et al. 1992, Numerical Recipes in Fortran 77. Cambridge University Press, London}

\bibitem{}
Radhakrishnan V. \& Srinivasan G. 1982, Curr. Science, 51, 1096

\bibitem{}
Romanova M. M., Kulkarni A. K. \& Lovelace R. V. E. 2008, ApJ., 673,
L171

\bibitem{}
Rappaport S., Verbunt F. \& Joss P. C. 1983, ApJ., 275, 713

\bibitem{}
Taam R. E. \& van den Heuvel E. P. J. 1986, ApJ., 305, 235

\bibitem{}
Urpin V., Geppert U. \& Konenkov D. 1998, A\&A,
331, 244
%On the origin of millisecond pulsars

\bibitem{}
van den Heuvel E. P. J., 2004, Science, 303, 1143

\bibitem{}
van den Heuvel, E. P. J., 2011,
Bulletin of the Astronomical Society of India,
39, 1
%Compact stars and the evolution of binary systems

\bibitem{}  van den Heuvel E. P. J. \& Bitzaraki O. 1995a, A\&A, 297, L41
%The magnetic field strength versus orbital period relation for binary radio
%pulsars with low-mass companions:
%evidence for neutron-star formation by accretion-induced collapse?

\bibitem{}
van den Heuvel E. P. J. \& Bitzaraki O. 1995b, In: The Lives of the
Neutron Stars, Kluwer Academic Publishers, Dordrecht

\bibitem{}
Wang J., Zhang C. M., Zhao Y. H., Kojima Y., Yin H. X., Song L. M.
2011, A\&A, 526, A88 %arXiv:1011.5013(astro-ph)

\bibitem{}
Wijnands R. \& van der Klis M. 1998, Nature, 394, 344
%A millisecond pulsar in an X-ray binary system

\bibitem{}
Wijers R. A. M. J. 1997, \mnras, 287, 607

\bibitem{}
Young E.J.  \& Chanmugam G.  1995, ApJ, 442, L53

\bibitem{}
Zhang C. M. \& Kojima Y. 2006, MNRAS, 366, 137

\bibitem{}
Zhang C. M., Wang J., Zhao Y. H., Yin H. X., Song L. M., Menezes D.
P., Wickramasinghe D. T., Ferrario L., Chardonnet P. 2011, A\&A,
527, 83

\end{thebibliography}
\end{document}